\documentclass[%
prd,
nofootinbib,
superscriptaddress
]{revtex4}
\usepackage{amsmath,amssymb,bm,float,mathrsfs}
\usepackage[dvipdfmx]{graphicx}
\usepackage{dcolumn}
\usepackage{color}
\usepackage{hyperref}
\usepackage{comment}
\usepackage{framed}
\usepackage{url}
\hypersetup{
    colorlinks=true,
    citecolor=cyan,
}

  \newcommand{\al}[1]{\begin{align} #1 \end{align}}
  
  \def\dd{\mathrm{d}}
  
  \def\pd{\partial}

\begin{document}

\title{Tensor clustering fossils in modified gravity and high-redshift gravitational-wave sound speed}


\author{Daisuke Yamauchi}
\email[Email: ]{d-yamauchi"at"ous.ac.jp}
\affiliation{
Department of Physics, Faculty of Science, Okayama University of Science, 1-1 Ridaicho, Okayama, 700-0005, Japan
}

\begin{abstract}
We investigate the tensor clustering fossils as a possible probe to constrain the theory of gravity, 
in particular the deviation of the sound speed of gravitational waves from the speed of light at high redshifts.
We develop the formalism of the effective Poisson equation to include the novel phenomenological model 
of the scalar-tensor tidal interactions
that are expected to be induced by the modification of the theory of gravity.
We show that the tensor clustering fossils can arise from the propagation of gravitational waves, 
the growth of the large-scale structures, and the second-order contributions from the effective Poisson equation. 
We construct the small-scale effective Lagrangian from the Horndeski scalar-tensor theory and derive
the formula applicable to the tensor clustering fossils in the language of the effective field theory of dark energy.
As a demonstration, we apply the formalism to the constraint on the sound speed of gravitational waves 
in the futuristic survey.
\end{abstract}


\maketitle

\section{Introduction}
\label{sec:Introduction}

The clustering of large-scale structure has been intensively studied using
the two-point correlation function in real space, or the power spectrum in Fourier space, 
which are determined under the assumption of statistical anisotropy.
When large-scale tensor perturbations (gravitational waves: GWs) couple to the scalar sector,
they tidally imprint local anisotropy in the small-scale distribution of matter and
the effect appears as local deviations from statistical anisotropy~\cite{Masui:2010cz,Jeong:2012df,Dai:2013kra,Schmidt:2013gwa,Masui:2017fzw}.
Interestingly, this imprint constitutes a fossilized map of the primordial tensor field 
since it persists even after the tensor mode has decayed by redshifting.
This fossil effect can arise from the coupling between the scalar and tensor perturbations not only 
during the inflationary epoch~\cite{Dimastrogiovanni:2014ina,Dimastrogiovanni:2015pla,Dimastrogiovanni:2019bfl}
but also during the post-inflationary epoch~\cite{Dai:2013kra}.
It can be shown that the detection of primordial GWs from the tensor clustering fossils 
in the standard cosmological model requires measurements with a huge dynamic range,
which is probably beyond the reach of current galaxy surveys, 
but such measurements would be achievable using the 21-cm line probes of neutral hydrogen during the dark ages~\cite{Masui:2010cz,Jeong:2012df}.
There are many projects to measure the 21-cm line in the dark ages from the far side of the Moon or from a satellite
orbiting around the Moon, such as FARSIDE~\cite{Burns:2021pkx}, DAPPER~\cite{Burns:2021ndk}, 
NCLE~\cite{2020AdSpR..65..856B}, LCRT~\cite{LCRT}, DSL~\cite{Chen:2019xvd}, TREED~\cite{TREED}, TSUKUYOMI~\cite{TSUKUYOMI} and so on.

In order to accurately evaluate the tensor clustering fossils, 
we must correctly take into account the nonlinear scalar-tensor tidal interactions induced during the late-time Universe.
Even if the initial density fluctuations are well described by linear theory, 
the nonlinearity of the gravitational dynamics eventually dominates the large-scale structure and the nonlinear mode coupling
is naturally induced even during the post-inflationary evolution,
since the continuity and Euler equations have nonlinear terms.
Furthermore, the modification of the underlying theory of gravity alters the nonlinear growth history of structure through the Poisson equation.
The quasi-nonlinear growth of large-scale structures can provide new insights into the theory of gravity
that would not be imprinted in the linear perturbation theory~\cite{Yamauchi:2017ibz,Yamauchi:2022fss,Yamashita:2024mdn,Sugiyama:2023tes,Sugiyama:2023zvd,Yamauchi:2021nxw,Arai:2022ilw}.
Although the detailed predictions depend on models of modified gravity, we can introduce phenomenological functions in the Poisson equation 
to characterize the deviations from the standard cosmological model.
The phenomenological model allows us to discuss general constraints on the modifications of gravity independently of the details of the gravity model.

One of the most stringent constraints on the theory of gravity is obtained from the nearly simultaneous detection of the GW
event GW170817 and its optical counterpart GRB170817A, which gave the constraint 
that the sound speed of GWs propagating from a neutron star binary should not deviate from 
that of light~\cite{TheLIGOScientific:2017qsa,Monitor:2017mdv,GBM:2017lvd}.
This observation contradicts any extension of general relativity that predicts 
a large deviation of the GW sound speed and 
implies that the sound speed of GWs is strictly coincident with the speed of light~\cite{Ezquiaga:2017ekz,Baker:2017hug,Sakstein:2017xjx,Creminelli:2017sry,Langlois:2017dyl}.
However, the propagation of GWs from GW170817 as well as the local test of gravity can place any constraints 
on the theory of gravity at the relatively low redshift, $z\lesssim 0.01$. 
There is still a large viability of theory of gravity whose deviation from general relativity becomes apparent 
at high redshifts (see e.g., \cite{TerenteDiaz:2023iqk,Lee:2022cyh}).

In this paper, we consider the tensor clustering fossils as a probe of the theory of gravity,
in particular the deviation of the GW sound speed from the speed of light at high redshifts.
To do this, we extend the formulation of the tensor clustering fossils to include the phenomenological model 
of the scalar-tensor tidal interactions in the effective Poisson equation. 
We then consider the specific theory of gravity, which naturally induces 
the deviation of the GW sound speed from the speed of light, 
and explore the nature of GWs at high redshifts by using the resulting formula of the tensor clustering fossils.

This paper is organized as follows. In Sec.~\ref{sec:General argument}, 
we derive the formula of the tensor clustering fossils,
which can be applied even in the situation that the nonlinear effective Poisson equation is modified.
In Sec.~\ref{sec:Tensor clustering fossils in modified gravity theory}, we compute the tensor clustering fossils
in the modified theory of gravity, in particular the Horndeski scalar-tensor theory. 
We construct the small-scale effective Lagrangian with the nonlinear mode-coupling between 
the scalar and tensor modes and show the explicit expression of the tensor clustering fossils induced
during the post-inflationary evolution.
We then qualitatively estimate the impact of the deviation of the GW speed from the speed of light 
by using the resulting formula of the tensor clustering fossils.
Finally, Sec.~\ref{sec:Conclustion} is devoted to the summary and discussion.

\section{Effective Poisson equation and tensor clustering fossils}
\label{sec:General argument}

In this section, we will derive the general expression for the tensor clustering fossils to include the effect
of the nonlinear mode-coupling between the scalar and tensor modes in the late-time Universe.
In particular, we extend the analysis of Ref.~\cite{Dai:2013kra} by incorporating the effect of the phenomenological model
in the effective Poisson equation.
Throughout this paper, we consider a metric with a Friedmann-Lema\^\i tre-Robertson-Walker (FLRW) background in the Newton gauge, 
which is given by
\al{
	\dd s^2=-\Bigl[1+2\Phi (t,{\bm x})\Bigr]\dd t^2
				+a^2(t)\biggl\{\Bigl[1-2\Psi(t,{\bm x})\Bigr]\delta_{ij}+h_{ij}(t,{\bm x})\biggr\}\dd x^i\dd x^j
	\,.\label{eq:Newton gauge metric}
}
Here, $h_{ij}$ denotes the transverse-traceless tensor mode.
We treat scalar and matter perturbations as independent small parameters from tensor perturbations, 
and we only keep their cross terms at second order.
Matter perturbations including the matter density contrast $\delta$ and peculiar velocity $v_i$ grow
from the primordial scalar perturbations. 
Assuming the pressure and anisotropic stress are neglected, and the matter is minimally coupled to gravity, 
the energy-momentum for the pressureless nonrelativistic matter is given by
\al{
	&T^0{}_0=-\rho_{\rm m}(t)\Bigl[1+\delta (t,{\bm x})\Bigr]
	\,,\\
	&T^0{}_i=a(t)\rho_{\rm m}(t)v_i(t,{\bm x})
	\,,\\
	&T^i{}_0=-\frac{1}{a(t)}\rho_{\rm m}(t)\Bigl[\delta^{ij}-h^{ij}(t,{\bm x})\Bigr] v_j(t,{\bm x})
	\,,\\
	&T^i{}_j=0
	\,.
}
To simplify the analysis, throughout this paper, we apply the quasi-static approximation. 
In particular, the wavenumbers of the scalar and tensor perturbations, which represent $k$
and $K$ respectively, are much larger than the comoving Hubble length scale.
The continuity and Euler equations for the pressureless nonrelativistic matter are given by
\al{
	&\dot\delta +\frac{1}{a}\left(\delta^{ij}-h^{ij}\right)\pd_iv_j =0
	\,,\label{eq:continuity}\\
	&\dot v_i+Hv_i+\frac{1}{a}\pd_i\Phi =0
	\,,\label{eq:Euler}
}
where the dot denotes the derivative with respect to the cosmic time $t$.
Although these fluid equations are the same as the standard ones in general relativity, there appears the effect of 
modification of the theory of gravity through the gravitational potential $\Phi$.
To obtain the closed-form equation for the matter perturbations, we need to add the information
for the relation between the gravitational potential and the matter density contrast, i.e., the Poisson equation.
In this paper, we formally consider the following form of the effective Poisson equation
\footnote{
The effective Poisson equation may include terms such as $\dot\delta$ and $\ddot\delta$.
When the Degenerate Higher-Order Scalar-Tensor (DHOST) theory of gravity~\cite{Langlois:2015cwa,Crisostomi:2016czh,Achour:2016rkg,BenAchour:2016fzp,Langlois:2017dyl} 
beyond the Horndeski theory~\cite{Horndeski:1974wa,Deffayet:2011gz,Kobayashi:2011nu}
is considered as gravity theory governing the structure formation, 
there can be shown to appear such time-derivatives of density contrast~\cite{Hirano:2019nkz,Hirano:2020dom}. 
In this paper, we simply neglect such terms for simplicity. 
}:
\al{
	\frac{1}{a^2H^2}\pd^2\Phi (t,{\bm x})
		=\frac{3}{2}\Omega_{\rm m}(t)\mu (t)\delta (t,{\bm x})
			+\left( -\Gamma_1(t)\frac{1}{a^2H^2}\pd^2h_{ij}(t,{\bm x})
					+\Gamma_2(t)\frac{1}{H}\dot h_{ij}(t,{\bm x})	\right)
			\frac{\pd_i\pd_j}{\pd^2}\delta (t,{\bm x})+\cdots
	\,,\label{eq:Poisson}
}
where $\Omega_{\rm m}$ denotes the energy fraction of the matter field, and 
we have introduced the time-dependent phenomenological functions $\Gamma_1$ and $\Gamma_2$,
which are related to the underlying gravity model.
When general relativity is considered, the second-order contribution to the effective Poisson equation \eqref{eq:Poisson} is given by 
$h^{ij}(\pd_i\pd_j/\pd^2)\delta$.
Since this term is suppressed by the factor $a^2H^2/K^2(\ll 1)$ compared with 
the terms appearing in Eq.~\eqref{eq:Poisson} in the subhorizon limit, we can drop such term.
In the subsequent analysis,
we will show such contribution can be safely neglected when one computes the tensor clustering fossils.
Therefore, in the case of the standard cosmology with general relativity, we take $\Gamma_1=\Gamma_2=0$.
The nonvanishing phenomenological functions $\Gamma_1$ and $\Gamma_2$ can be treated as the signals 
of the deviation from general relativity.

In this paper, we will solve these equations of motion by using the perturbed expansion defined as
\al{
	\delta =\sum_{n=1}\delta^{(n)}
	\,,\ \ \ 
	\Phi =\sum_{n=1}\Phi^{(n)}
	\,,\ \ \ 
	\cdots\,,
}
where $\delta^{(n)}$\,, $\Phi^{(n)}$\,, $\cdots ={\cal O}(\epsilon^n )$ are the $n$-th order quantities with $\delta^{(1)}$
being ${\cal O}(\epsilon)$ quantity.
At linear order, the closed-form equation for the density contrast is given by
\al{
	\ddot\delta^{(1)} +2H\dot\delta^{(1)} -\frac{3}{2}H^2\Omega_{\rm m}\mu\delta^{(1)} =0
	\,,\label{eq:1st order eq}
}
When the time dependence of the effective gravitational coupling $\mu$ is given,
one can numerically solve Eq.~\eqref{eq:1st order eq}.
In general, Eq.~\eqref{eq:1st order eq} has two independent solutions, which we call $D_\pm (t)$.
We take the boundary condition of the growing mode as
$\delta^{(1)}(t,{\bm x})=D_+ (t)\delta_{\rm prim}({\bm x})$, where
$\delta_{\rm prim}({\bm x})$ denotes the primordial density contrast defined at the very early epoch.
The linear peculiar velocity field $v_i^{(1)}$ can  be described in terms of the linear density contrast through Eq.~\eqref{eq:continuity}
as $v_i^{(1)}=-(\pd_i/\pd^2)\dot\delta^{(1)}=-Hf(\pd_i/\pd^2)\delta^{(1)}$,
where $f=\dd\ln D_+/\dd\ln a$ denotes the growth rate of the density contrast. 
On the other hand, as for the tensor mode, we assume the form of the evolution equation as
\al{
	\ddot h_{ij}+\left(3+\nu\right) H\dot h_{ij}-c_{\rm T}^2\frac{\pd^2}{a^2}h_{ij}=0
	\,,\label{eq:GW EoM}
}
where $\nu$ and $c_{\rm T}^2$ represent the time-dependent phenomenological functions characterizing
the properties of the large-scale tensor mode.

Let us consider the second-order density contrast induced by the scalar-tensor tidal interactions.
Substituting the first-order solutions of the scalar and tensor modes 
into Eqs.~\eqref{eq:continuity}--\eqref{eq:Euler} as the source terms, 
we obtain the second-order equation as
\al{
	\ddot\delta^{(2)}+2H\dot\delta^{(2)}-\frac{\pd^2}{a^2}\Phi^{(2)}
		=-Hf\dot h^{ij}\frac{\pd_i\pd_j}{\pd^2}\delta^{(1)}+\frac{3}{2}H^2\Omega_{\rm m}\mu\, h^{ij}\frac{\pd_i\pd_j}{\pd^2}\delta^{(1)}
	\,.\label{eq:2nd order eq2}
}
Since the first and second terms in the right hand side of the above equation are estimated as 
${\cal O}(H\dot h\delta )\sim {\cal O}((K^2/a^2) h\delta )$ and ${\cal O}(H^2h\delta )$ respectively,
the first term is ${\cal O}(K^2/a^2H^2)$ larger than the second term in the subhorizon limit.
We therefore neglect the second term in the right hand side of Eq.~\eqref{eq:2nd order eq2}. 
We then use the second-order effective Poisson equation \eqref{eq:Poisson} to obtain
the closed-form equation for the second-order density contrast as
\al{
	\ddot\delta^{(2)}+2H\dot\delta^{(2)}-\frac{3}{2}H^2\Omega_{\rm m}\mu\delta^{(2)}
		=\Bigl\{
			-\Gamma_1\frac{1}{a^2}\pd^2 h^{ij}
			-\left( f-\Gamma_2\right)H\dot h^{ij}\,
		\Bigr\}\frac{\pd_i\pd_j}{\pd^2}\delta^{(1)}
	\,.
}
Introducing the following quantity:
\al{
	\widetilde\delta^{(2)}=\delta^{(2)}-h^{ij}\frac{\pd_i\pd_j}{\pd^2}\delta^{(1)}
	\,,
}
and using Eq.~\eqref{eq:GW EoM}, we can rewrite Eq.~\eqref{eq:2nd order eq} as
\al{
	\ddot{\widetilde\delta}{}^{(2)}+2H\dot{\widetilde\delta}{}^{(2)}-\frac{3}{2}H^2\Omega_{\rm m}\mu\,\widetilde\delta^{(2)}
		=\Bigl\{
			\left(c_{\rm T}^2-\Gamma_1\right)\frac{1}{a^2}\pd^2 h^{ij}
			+\left( f-1-\nu +\Gamma_2\right)H\dot h^{ij}\,
		\Bigr\}\frac{\pd_i\pd_j}{\pd^2}\delta^{(1)}
	\,,\label{eq:2nd order eq}
}
where we have dropped the sub-leading terms.
The solution of the above equation can be divided into two pieces:
$\widetilde\delta^{(2)}(t ,{\bm x})=\delta^{(2)}_{\rm homo}(t ,{\bm x})+\delta^{(2)}_{\rm spec}(t ,{\bm x})$, 
where $\delta^{(2)}_{\rm homo}$ denotes the solution of the homogeneous part of Eq.~\eqref{eq:2nd order eq},
and $\delta^{(2)}_{\rm spec}$ is a special solution that solves the full equation with the vanishing boundary condition at 
the very early epoch.
The special solution can be constructed by using the Green function method.
Using the conformal time defined as $\eta=\int^t\dd \bar t/a(\bar t)$ and transforming to Fourier space, 
we obtain the special solution of the form
\al{
	\delta_{\rm spec}^{(2)}(\eta ,{\bm k})
		=-h_{ij}^{\rm prim}({\bm K})\widehat k^i\widehat k^j\delta^{(1)}(\eta ,{\bm k})
			{\cal S}(\eta ,K)
	\,,
}
where the prime denotes the derivative with respect to the conformal time, $\widehat k^i\equiv k^i/k$, and 
${\cal S}$ is given by
\al{
	{\cal S}(\eta ,K)
		=\int_0^\eta\dd\bar\eta
			\biggl\{
				\Bigl[ c_{\rm T}^2(\bar\eta)-\Gamma_1(\bar\eta) \Bigr]K^2{\cal T}_h(\bar\eta ,K)
				-\Bigl[f(\bar\eta )-1-\nu(\bar\eta )-\Gamma_2(\bar\eta)\Bigr]{\cal H}(\bar\eta){\cal T}_h^\prime (\bar\eta ,K)
			\biggr\}
			\frac{D_+(\bar\eta)}{D_+(\eta )}
			G_{\rm ret}(\eta ,\bar\eta )
	\,,\label{eq:calS}
}
with ${\cal H}=aH$.
We have introduced the transfer function of the tensor mode as 
$h_{ij}(\eta ,{\bm K})={\cal T}_h(\eta ,K)h_{ij}^{\rm prim}({\bm K})$ with $h_{ij}^{\rm prim}({\bm K})$ being
the primordial tensor mode.
Here, $G_{\rm ret}$ denotes the retarded Green function 
which can be written in terms of the two independent solutions of the homogeneous part as
\al{
	G_{\rm ret}(\eta ,\bar\eta )
		=\frac{D_+(\eta)D_-(\bar\eta)-D_+(\bar\eta)D_-(\eta)}{D_+^\prime(\bar\eta)D_-(\bar\eta )-D_+(\bar\eta )D_-^\prime(\bar\eta )}
			\Theta (\eta -\bar\eta)
	\,,
}
where $\Theta (x)$ denotes the step function.
Combining the resultant expressions derived in this section, 
the solution of the matter density contrast valid up to the second order can be expressed as
\al{
	\delta (\eta, {\bm k})=\delta^{(1)}(\eta ,{\bm k})
			-h_{ij}^{\rm prim}({\bm K})\widehat k^i\widehat k^j
				\biggl[\frac{1}{2}\frac{\dd\ln{\cal T}_\delta}{\dd\ln k}{\cal T}_h(\eta ,K)+{\cal S}(\eta ,K)\biggr]\delta^{(1)}(\eta ,{\bm k})
	\,.\label{eq:standard result}
}
where we have introduced the linear transfer function for the density contrast, defined as $\delta^{(1)}(\eta ,{\bm k})={\cal T}_\delta (\eta, k)\Phi_{\rm prim}({\bm k})$.
In our prescription, $\dd\ln{\cal T}_\delta /\dd\ln k=2$ in the subhorizon limit.
As we will show later, even in the case of the modified gravity, ${\cal S}$ is expected to persist after the tensor mode has decayed.
Therefore, we still refer to this effect as tensor clustering fossils.
We find from Eq.~\eqref{eq:calS} that the tensor clustering fossils can generally be induced by three parts:
the properties of the large-scale GWs ($c_{\rm T}^2$ and $\nu$), 
the deviation of the structure growth rate from unity ($f-1$), and the second-order scalar-tensor tidal interactions 
from the effective Poisson equations ($\Gamma_1$ and $\Gamma_2$).
These effects are expected to increase the amplitude of the tensor clustering fossils.
Even when the contributions from $(f-1)$ and $\Gamma_{1,2}$ can be neglected,
the modification of the tensor clustering fossils still exists.
In other words, we can use the tensor clustering fossils \eqref{eq:calS} to constrain 
the properties of large-scale GWs. 
One can easily find that if we consider general relativity and matter-dominated era, namely 
$D_+\propto\eta^2$\,, $D_-\propto 1/\eta^3$\,, $c_{\rm T}^2=1$\,, and
$\nu =\Gamma_1=\Gamma_2=0$, Eq.~\eqref{eq:calS} reduces to the standard form 
of the tensor clustering fossils, which was previously derived in Ref.~\cite{Dai:2013kra}.

We further assume that the primordial fields are generated through the standard inflationary scenario, 
in which the primordial scalar-scalar-tensor bispectrum satisfies the consistency relation in the squeezed limit.
In the limit of $K\ll k$, the primordial correlations between large-scale tensor and small-scale scalar modes 
lead to an anisotropic primordial scalar power spectrum as~\cite{Giddings:2010nc,Giddings:2011zd}
\al{
	\widetilde P_\Phi ({\bm k})\Bigl|_{h_{ij}({\bm K})}=P_\Phi (k)\biggl[1-\frac{1}{2}h_{ij}^{\rm prim}\widehat k^i\widehat k^j\frac{\dd\ln P_\Phi}{\dd\ln k}\biggr]
	\,.
}
Here, $|_{h_{ij}({\bm K})}$ represents an ensemble average over realizations of
the scalar perturbations while holding the tensor perturbations constant.
We then obtain the anisotropic power spectrum 
for matter perturbations~\cite{Masui:2010cz,Jeong:2012df,Dai:2013kra,Schmidt:2013gwa,Masui:2017fzw}:
\al{
	\widetilde P(\eta,{\bm k})\Bigl|_{h_{ij}({\bm K})}
		=P (\eta, k)\biggl\{ 1-h_{ij}^{\rm prim}({\bm K})\widehat k^i\widehat k^j
			\biggl[\frac{1}{2}\frac{\dd\ln P}{\dd\ln k}+2{\cal S}_N(\eta, K)\biggr]\biggr\}
		\,,
}
where
\al{
	{\cal S}_N(\eta ,K)={\cal S}(\eta ,K)-\frac{1}{2}\frac{\dd\ln{\cal T}_\delta}{\dd\ln k}\Bigl[1-{\cal T}_h(\eta ,K)\Bigr]
	\,.
}
Once the underlying gravity model is specified, we can calculate the tensor clustering fossils by using the formula derived above.
In the subsequent analysis, we focus on the Horndeski class of modified gravity which naturally induces the deviation
of the sound speed of GWs from the speed of light. 

\section{Tensor clustering fossils in modified gravity}
\label{sec:Tensor clustering fossils in modified gravity theory}

\subsection{Horndeski scalar-tensor theory}

In this section, let us consider the Horndeski scalar-tensor theory~\cite{Horndeski:1974wa,Deffayet:2011gz,Kobayashi:2011nu}
(see \cite{Langlois:2018dxi,Kobayashi:2019hrl,Arai:2022ilw} for a review) as 
the theory of gravity governing the clustering of the matter density contrast.
The Horndeski theory is a most general scalar-tensor gravity theory with the second-order equations-of-motion,
to which attention is paid as one of the attractive modified gravity theories.
The Lagrangian of the Horndeski theory is given by
\al{
	{\cal L}_{\rm grav}
		=\sqrt{-g}\biggl\{
			G_2(\phi ,X)-G_3(\phi ,X)\Box\phi +G_4(\phi ,X)R
			+G_{4X}\Bigl[(\Box\phi)^2-\nabla^\mu\nabla^\nu\phi\nabla_\mu\nabla_\nu\phi\Bigr]
		\biggr\}
	\,,\label{eq:Horndeski Lag}
}
where $G_2$\,, $G_3$\,, and $G_4$ are arbitrary functions of the scalar field $\phi$ and
its kinetic term $X:=-g^{\mu\nu}\nabla_\mu\phi\nabla_\nu\phi /2$, and 
we have introduced the notation: $g_\phi =\partial g/\partial\phi$, $g_X:=\partial g/\partial X$, and so on.
A number of scalar-tensor theories such as
the Jordan-Brans-Dicke theory~\cite{Brans:1961sx}, the $f(R)$ gravity~\cite{Starobinsky:1980te}, 
the $k$-essence~\cite{Chiba:1999ka}, the kinetic gravity braiding~\cite{Deffayet:2010qz}, and 
the non-minimal coupling to the Gauss-Bonnet term~\cite{Kobayashi:2011nu} belong to the Horndeski family.
In particular, the Horndeski theory naturally includes the models with the deviation of the GW
sound speed from the speed of light. 
We note that we have simply dropped the ``$G_5$'' terms in Eq.~\eqref{eq:Horndeski Lag},
but the generic features presented in the subsequent analysis are expected to be the same, 
and a wider class of modified gravity such as
DHOST theory~\cite{Langlois:2015cwa,Crisostomi:2016czh,Achour:2016rkg,BenAchour:2016fzp,Langlois:2017dyl} 
would be straightforward.

To describe cosmological perturbations around the FLRW background in theories with
a preferred slicing induced by a time-dependent scalar field, it is convenient to use
the language of the effective field theory (EFT) of dark energy.
In this context, instead of considering the explicit form of the Horndeski functions $G_a$,
we will introduce the EFT parameters to determine completely the dynamics of cosmological perturbations.
Moreover, we focus only on the operators that contribute in the quasi-static limit.
In this limit, the minimum set to specify the total amount of cosmological information up to 
the linear order is a set of three independent functions of time that can be labelled by
$\{\alpha_{\rm T},\alpha_{\rm M},\alpha_{\rm B}\}$ in addition to the Hubble parameter
$H$ and the effective Planck mass $M$. 
The EFT parameters are related explicitly to the Horndeski functions through~\cite{Bellini:2014fua}
\al{
	&M^2=2G_4-4XG_{4X}
	\,,\\
	&MH^2\alpha_{\rm M}=\frac{\dd}{\dd t}M^2
	\,,\\
	&HM^2\alpha_{\rm B}=2\dot\phi\left( XG_{3X}-G_{4\phi}-2XG_{4\phi X}\right) +8HX\left( G_{4X}+2XG_{4XX}\right)
	\,,\\
	&M^2\alpha_{\rm T}=4XG_{4X}
	\,.
}
To generalize this formalism to the second-order perturbations in the subhorizon scales,
one other function of time is required~\cite{Yamauchi:2017ibz,Bellini:2015wfa}. The explicit form is given by
\al{
	M^2\alpha_{\rm V}=2X\left(G_{4X}+2XG_{4XX}\right)
	\,.
}
Here, the right hand sides of these EFT parameters are evaluated at the cosmological background.

For the purpose of studying the clustering fossils in the modified gravity, 
we will derive the small-scale effective Lagrangian for the scalar and tensor modes, following 
Refs.~\cite{Kobayashi:2014ida,Koyama:2013paa,Hirano:2020dom,Yamauchi:2021nxw}.
Without loss of generality, we take $\phi (t,{\bm x})=t+\pi (t,{\bm x})$.
We now expand the full Horndeski Lagrangian Eq.~\eqref{eq:Horndeski Lag} around 
the FLRW background to obtain the effective Lagrangian using the following assumptions:
As for the scalar perturbations, we employ the quasi-static approximation so that 
$(\pd\epsilon)^2\gg (\dot\epsilon)^2$ in the Lagrangian,
where $\epsilon$ stands for any of $\Phi$, $\Psi$\,, and $H\pi$.
On the other hand, as for the tensor perturbations, its time derivative provides
the same order of magnitude as its spatial derivative.
Hence, we will keep the scalar perturbations with the highest spatial derivatives and 
the tensor perturbations with all the time and spatial derivatives.
Moreover, in this paper, we only take into account the scalar-scalar-tensor 
three-point interaction terms with the highest derivatives.
By doing this expansion, we schematically find the following terms:
\al{
	(\pd\epsilon)^2\,,\ \ (\pd h)^2\,,\ \ \dot h^2\,,\ \ 
	\pd^2 h(\pd\epsilon)^2\,\ \ \dot h(\pd\epsilon)^2\,,\ \ \ddot h(\pd\epsilon)^2
	\,.
}
By explicitly expanding the full Horndeski Lagrangian Eq.~\eqref{eq:Horndeski Lag} under
these prescriptions, we find the small-scale effective Lagrangian valid up to the third order as
\al{
	{\cal L}_{\rm eff}
		=&\frac{M^2a}{2}
			\biggl[
				4\Psi\pd^2\Phi
				-2\left( 1+\alpha_{\rm T}\right)\Psi\pd^2\Psi
				-a^2(\dot h_{ij})^2+\left( 1+\alpha_{\rm T}\right)(\pd_kh_{ij})^2
	\notag\\
	&\quad\quad
			+4H\left(\alpha_{\rm M}-\alpha_{\rm T}\right)\Psi\pd^2\pi
			-4H\alpha_{\rm B}\Phi\pd^2\pi
			+H^2c_{\pi\pi}\pi\pd^2\pi
	\notag\\
	&\quad\quad
			+2\alpha_{\rm T}\dot h^{ij}\Psi\pd_i\pd_j\pi
			-4\alpha_{\rm V}\dot h^{ij}\Phi\pd_i\pd_j\pi
	\notag\\
	&\quad\quad
			-\frac{5}{a^2}\alpha_{\rm T}\pd^2h^{ij}\pi\pd_i\pd_j\pi
			+Hc_{\pi\pi\dot h}\dot h^{ij}\pi\pd_i\pd_j\pi
			-2\alpha_{\rm V}\ddot h^{ij}\pi\pd_i\pd_j\pi
			\biggr]
	\,,\label{eq:eff Lag}
}
where the coefficients $c_{\pi\pi}$ and $c_{\pi\pi\dot h}$ are defined as
\al{
    &c_{\pi\pi} = 
        -2\biggl\{\,
		\frac{\dot H}{H^2}+\frac{3}{2}\Omega_{\rm m}
		+\alpha_{\rm T}-\alpha_{\rm M}
		+\frac{(aM^2H\alpha_{\rm B})^\cdot}{aM^2H^2}
	\biggr\}
    \,,\\
    &c_{\pi\pi\dot h}=2\biggl\{
				\left(\alpha_{\rm T}-\alpha_{\rm M}-\alpha_{\rm B}\right)
				-\frac{(aM^2\alpha_{\rm V})^\cdot}{aM^2H}\biggr\}
	\,.
}
Based on this effective Lagrangian, we then study the quasi-static behavior of those perturbations deep inside the cosmological horizon.
Varying the Lagrangian Eq.~\eqref{eq:eff Lag} with respect to $\Phi$, $\Psi$, and $\pi$,
we derive the first-order equations of motion for the scalar perturbations as  
\al{
	&\pd^2\bigl(\Psi^{(1)} -\alpha_{\rm B}H\pi^{(1)}\bigr) -\frac{3}{2}a^2H^2\Omega_{\rm m}\delta^{(1)}=0
	\,,\\
	&\pd^2\Bigl[\Phi^{(1)} -(1+\alpha_{\rm T})\Psi^{(1)} +(\alpha_{\rm M}-\alpha_{\rm T})H\pi^{(1)}\Bigr]=0
	\,,\\
	&\pd^2\biggl[ (\alpha_{\rm M}-\alpha_{\rm T})\Psi^{(1)}-\alpha_{\rm B}\Phi^{(1)}+\frac{1}{2}c_{\pi\pi}H\pi^{(1)}\biggr]
		=0
	\,,
}
which can be easily solved analytically to obtain the following form:
\al{
	\pd^2\epsilon^{(1)} =\frac{3}{2}a^2H^2\Omega_{\rm m}\mu_\epsilon\delta^{(1)}
	\,,\label{eq:eps1 sol}
}
where $\epsilon =\Phi$\,, $\Psi$\,, $H\pi$\,, and their coefficients are given by (see also \cite{Pogosian:2016pwr,Gleyzes:2015rua,Hirano:2019nkz})
\al{
	&\mu_\Phi =1+\alpha_{\rm T}+\frac{2\xi^2}{c_{\rm s}^2\alpha}
	\,,\label{eq:Phi1 sol}\\
	&\mu_\Psi =1+\frac{2\alpha_{\rm B}\xi}{c_{\rm s}^2\alpha}
	\,,\\
	&\mu_{H\pi}=\frac{2\xi}{c_{\rm s}^2\alpha}
	\,,\label{eq:pi1 sol}
}
with $\xi=\alpha_{\rm B}(1+\alpha_{\rm T})+\alpha_{\rm T}-\alpha_{\rm M}$.
Here, we have introduced the following quantity:
\al{
	c_{\rm s}^2\alpha
		=&-2\biggl\{(1+\alpha_{\rm B})\biggl[\frac{\dot H}{H^2}+\alpha_{\rm B}(1+\alpha_{\rm T})+\alpha_{\rm T}-\alpha_{\rm M}\biggr]+\frac{\dot\alpha_{\rm B}}{H}+\frac{3}{2}\Omega_{\rm m}\biggr\}
	\,,\label{c_s^2alpha}
}
with $c_{\rm s}^2$ representing the propagation speed of the scalar mode and $\alpha >0$.
We note that $\alpha$ is also written in terms of other EFT parameters, but we don't show it explicitly.
For the tensor perturbations, the equation of motion can be derived by varying the effective Lagrangian with respect to $h_{ij}$:
\al{
	\ddot h_{ij}+\left(3+\alpha_{\rm M}\right) H\dot h_{ij}-\left(1+\alpha_{\rm T}\right)\frac{\pd^2}{a^2}h_{ij}=0
	\,,
}
which corresponds to $\nu =\alpha_{\rm M}$ and $c_{\rm T}^2=1+\alpha_{\rm T}$ in the context of the phenomenological model Eq.~\eqref{eq:GW EoM}.

We next consider the second-order perturbations induced by 
the scalar-tensor tidal interactions.
To obtain the second-order solutions of the gravitational potentials, 
we need to substitute the first-order solutions Eq.~\eqref{eq:eps1 sol}
into the interaction terms between the scalar and tensor modes derived
from Eq.~\eqref{eq:eff Lag}.
The second-order equations of motion are given by
\al{
	&\frac{2}{3a^2H^2\Omega_{\rm m}}\pd^2\bigl(\Psi^{(2)} -\alpha_{\rm B}H\pi^{(2)}\bigr) -\delta^{(2)}
		=\Pi_\Phi\frac{1}{H}\dot h^{ij}\frac{\pd_i\pd_j}{\pd^2}\delta^{(1)}
	\,,\\
	&\frac{2}{3a^2H^2\Omega_{\rm m}}\pd^2\Bigl[\Phi^{(2)} -(1+\alpha_{\rm T})\Psi^{(2)} +(\alpha_{\rm M}-\alpha_{\rm T})H\pi^{(2)}\Bigr]
		=\Pi_\Psi\frac{1}{H}\dot h^{ij}\frac{\pd_i\pd_j}{\pd^2}\delta^{(1)}
	\,,
}
for the gravitational potentials,
and
\al{
	\frac{2}{3a^2H^2\Omega_{\rm m}}\pd^2\biggl[ (\alpha_{\rm M}-\alpha_{\rm T})\Psi^{(2)}-\alpha_{\rm B}\Phi^{(2)}+\frac{1}{2}c_{\pi\pi}H\pi^{(2)}\biggr]
		=\biggl[-\Pi_{\pi ,1}\frac{1}{a^2H^2}\pd^2 h^{ij}+\Pi_{\pi ,2}\frac{1}{H}\dot h^{ij}\biggr]\frac{\pd_i\pd_j}{\pd^2}\delta^{(1)}
	\,,
}
for the scalar field perturbations.
The coefficients for the interaction terms are 
\al{
	&\Pi_\Phi =\alpha_{\rm V}\mu_{H\pi}
	\,,\label{eq:Pi_Phi}\\
	&\Pi_\Psi =-\frac{1}{2}\alpha_{\rm T}\mu_{H\pi}
	\,,\\
	&\Pi_{\pi ,1}=-\frac{1}{2}\Bigl\{5\alpha_{\rm T}+\alpha_{\rm V}\left(1+\alpha_{\rm T}\right)\Bigr\}\mu_{H\pi}
	\,,\\
	&\Pi_{\pi ,2}=\alpha_{\rm V}\mu_\Phi -\frac{1}{2}\alpha_{\rm T}\mu_\Psi 
				-\biggl\{ \frac{1}{2}c_{\pi\pi\dot h}+\alpha_{\rm V}\left(3+\alpha_{\rm M}\right)\biggr\}\mu_{H\pi}
	\,.\label{eq:Pi_pi2}
}
Combining these equations, we finally obtain the solution of the second-order gravitational potential induced by
the scalar-tensor tidal interactions as
\al{
	&\frac{1}{a^2H^2}\pd^2\Phi^{(2)}
		=\frac{3}{2}\Omega_{\rm m}\mu_\Phi\delta^{(2)}
			+\frac{3}{2}\Omega_{\rm m}\biggl[-\mu_{H\pi}\Pi_{\pi ,1}\frac{1}{a^2H^2}\pd^2 h^{ij}
				+\Bigl\{\mu_\Phi\Pi_\Phi+\kappa_\Psi\Pi_\Psi +\mu_{H\pi}\Pi_{\pi ,2}\Bigr\}\frac{1}{H}\dot h^{ij}
				\biggr]\frac{\pd_i\pd_j}{\pd^2}\delta^{(1)}
	\,.
}
In summary, the phenomenological functions $\mu$\,, $\Gamma_1$\,, and $\Gamma_2$, which were defined
in the effective Poisson equation \eqref{eq:Poisson}, can be written in terms of the EFT parameters as
\al{
	&\mu =\mu_\Phi
		=1+\alpha_{\rm T}+\frac{2[\alpha_{\rm B}(1+\alpha_{\rm T})+\alpha_{\rm T}-\alpha_{\rm M}]^2}{c_{\rm s}^2\alpha}
	\,,\label{eq:mu}\\
	&\Gamma_1
		=\frac{3}{2}\Omega_{\rm m}\,\mu_{H\pi}\Pi_{\pi ,1}
	\,,\label{eq:Gamma1}\\
	&\Gamma_2
		=\frac{3}{2}\Omega_{\rm m}\Bigl\{\mu_\Phi\Pi_\Phi+\mu_\Psi\Pi_\Psi +\mu_{H\pi}\Pi_{\pi ,2}\Bigr\}	
	\,,\label{eq:Gamma2}
}
where $\mu_i$ and $\Pi_i$ were defined in Eqs.~\eqref{eq:Phi1 sol}--\eqref{eq:pi1 sol} and \eqref{eq:Pi_Phi}--\eqref{eq:Pi_pi2}.
These expressions are one of the main results of this paper.
Once the time dependence of the EFT parameters is explicitly given, 
we can compute the coefficients of the effective Poisson equation
with the use of the above expressions and 
the tensor clustering fossils by using the formula Eq.~\eqref{eq:calS} derived in
the previous section.

\subsection{$\alpha_{\rm T}$ model}

In this subsection, we apply the resulting formula derived in the previous subsection
to the constraint on a specific parameter. 
In particular, we focus on the deviation of the GW sound speed from the speed of light. 
To do this, let us consider the specific setup such that all the EFT parameters except for $\alpha_{\rm T}$ are taken to be zero
and the Universe is dominated by the matter, namely $\alpha_{\rm M}=\alpha_{\rm B}=\alpha_{\rm V}=0$ and $\Omega_{\rm m}=1$.
We call this parametrization ``$\alpha_{\rm T}$ model''.
Under these assumptions, one finds that the phenomenological functions characterizing
the first-order effective Poisson equation [Eqs.~\eqref{eq:Phi1 sol}--\eqref{eq:pi1 sol}] in the $\alpha_{\rm T}$ model reduce to
\al{
	\mu_\Phi =\mu_\Psi =-\mu_{H\pi}=1
	\,.
}
This fact implies that the first-order equations for the scalar perturbations are the same
as the standard ones.
Hence, the gravitational potentials coincide with each other and these are conserved 
through matter domination, i.e.,
$\Phi^{(1)}=\Psi^{(1)}=\Phi_{\rm prim}$, and
the matter density contrast grows in proportion to the scale factor, $D_+(t)\propto a(t)$.

To discuss the effect of $\alpha_{\rm T}$ in the context of the tensor clustering fossils, we consider 
the second-order phenomenological functions given in Eqs.~\eqref{eq:Gamma1} and \eqref{eq:Gamma2}.
In the $\alpha_{\rm T}$ model, these functions can be easily shown to be
\al{
	&
	\Gamma_1=-\frac{15}{4}\alpha_{\rm T}
	\,,\ \ \ 
	\Gamma_2=0
	\,.
}
Substituting these into the general formula of the tensor fossils Eq.~\eqref{eq:calS}, we have
\al{
	{\cal S}(\eta ,K)
		=\frac{1}{5}\int_0^\eta\dd\bar\eta
			\left( 1+\frac{19}{4}\alpha_{\rm T}(\bar\eta )\right)
				K^2\bar\eta\,{\cal T}_h(\bar\eta ,K;\alpha_{\rm T})
				\Biggl[ 1-\left(\frac{\bar\eta}{\eta}\right)^5\,\Biggr]
	\,.\label{eq:reduced calS}
}
This expression provides the generic result in the $\alpha_{\rm T}$ model.
The time-dependent parameter $\alpha_{\rm T}(\eta )$ affects not only the coefficient but also
the transfer function of the tensor mode. 
As we will show, the modification of the transfer function provides non-negligible contributions
to the tensor clustering fossils.
We note that the value of $\alpha_{\rm T}$ should be chosen in the range of $-1\leq\alpha_{\rm T}\leq 0$ 
to avoid instability in the perturbation for the scalar and tensor modes.

%
\begin{figure}
\includegraphics[width=130mm]{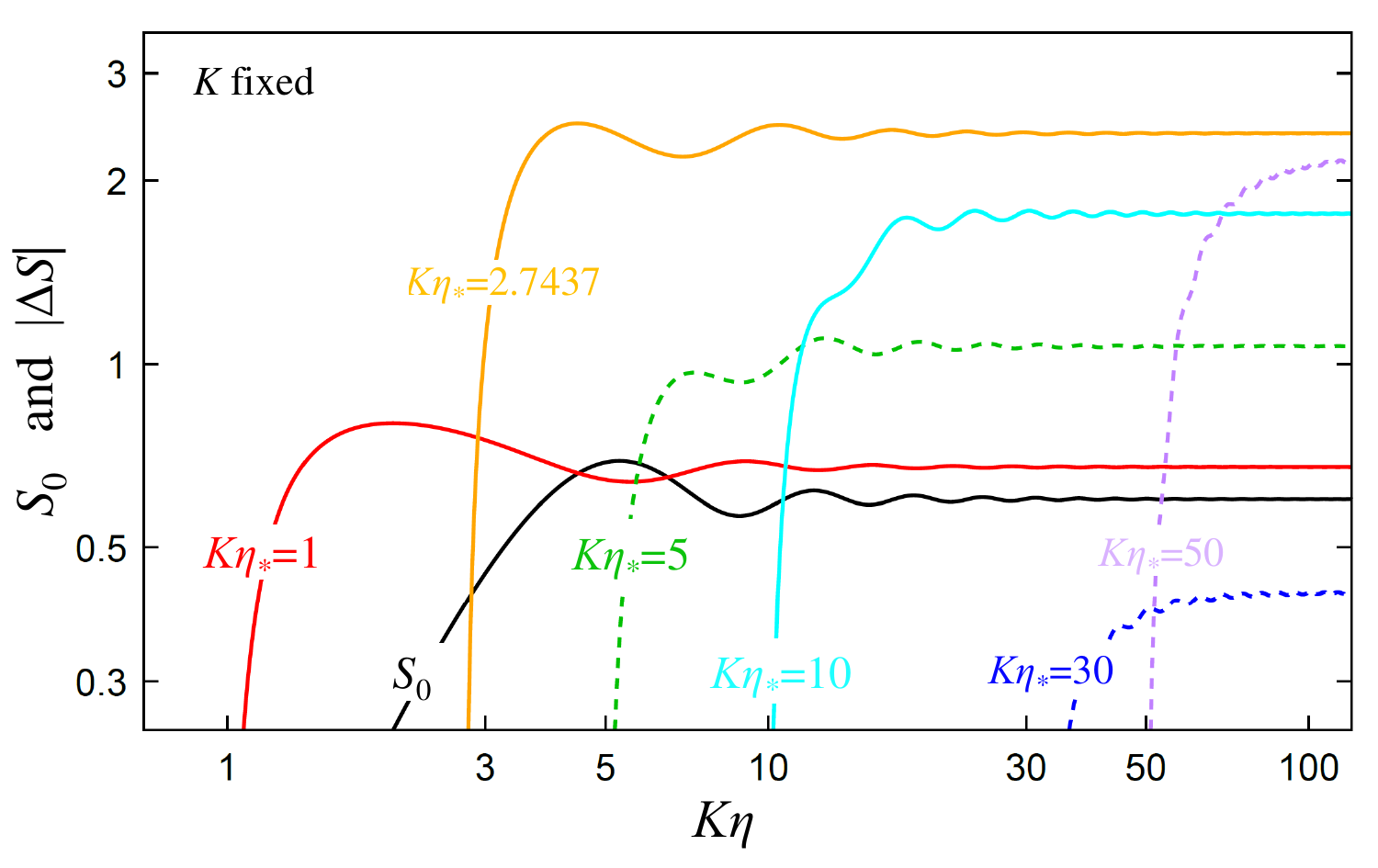}
\vspace{-2mm}
\caption{
The amplitude of the tensor clustering fossils ${\cal S}$ as a function of $K\eta$
while keeping $K\eta_\ast$ constant in the $\alpha_{\rm T}$ model.
We show the standard one ${\cal S}_0$ (black) and the correction factor $|\Delta{\cal S}|$ 
with $K\eta_\ast =1$ (red), $2.7437$ (orange), $5$ (green), $10$ (blue), $30$ (purple), and 
$50$ (magenta).
The solid and dashed curves represent the positive and negative values, respectively. 
}
\label{fig:S}
\end{figure} 
%

To continue the analysis, we now consider the specific time-dependence of $\alpha_{\rm T}$.
As mentioned in Sec.~\ref{sec:Introduction}, the observation of GW170817~\cite{TheLIGOScientific:2017qsa,Monitor:2017mdv,GBM:2017lvd} 
as well as the local test of gravity put a strong constraint on the GW sound speed at the relatively low redshift.
There is still a large viability of the theory of gravity where the deviation of the GW sound speed from the speed of light
appears only at high redshifts.
With this in mind, we focus on the case where the deviation of the GW sound speed from the speed of light
occurs only for a very short period during the (high-redshift) matter-dominated phase.
Specifically, we consider the following form of $\alpha_{\rm T}$:
\al{
	\alpha_{\rm T}(\eta )=\alpha_{\rm T}^\ast\,\delta_{\rm D} (\ln\eta -\ln\eta_\ast )
	\,.
}
Hereafter, we treat two constant model parameters $\alpha_{\rm T}^\ast$ and $\eta_\ast$ as the characteristic amplitude and time of the modification.
In this setup, the equation of motion for the tensor mode in the Fourier space is given by
\al{
	h_{ij}^{\prime\prime}+\frac{4}{\eta}h_{ij}^\prime +\Bigl[1+\alpha_{\rm T}^\ast\,\delta_{\rm D}(\ln\eta -\ln\eta_\ast )\Bigr]K^2h_{ij}=0
	\,.\label{eq:h EoM}
}
Since for $\eta <\eta_\ast$ there are no modifications, 
the transfer function of the tensor mode
is described by the standard one, that is ${\cal T}_h(K\eta )=3j_1(K\eta )/K\eta =:{\cal T}_{h,0}(K\eta)$, with $j_n(x)$ and $y_n(x)$ being 
the spherical Bessel functions of the first and second kinds.
To obtain the solution at $\eta >\eta_\ast$, we need to solve the junction conditions at $\eta =\eta_\ast$.
We then find the transfer function of the tensor mode valid even after $\eta >\eta_\ast$ as
\al{
	{\cal T}_h(\eta ,K)
		={\cal T}_{h,0}(K\eta )+\alpha_{\rm T}^\ast\,\Delta{\cal T}_h(\eta ,K)\,\Theta (\eta -\eta_\ast)
	\,,\label{eq:modified T_h}
}
with
\al{
	\Delta{\cal T}_h(\eta ,K)
		=3(K\eta_\ast)^3j_1(K\eta_\ast )
				\left(y_1(K\eta_\ast)\frac{j_1(K\eta)}{K\eta}- j_1(K\eta_\ast)\frac{y_1(K\eta)}{K\eta}\right)
	\,.
}
Substituting the transfer function Eq.~\eqref{eq:modified T_h} into Eq.~\eqref{eq:reduced calS}, 
the tensor clustering fossils with the instantaneous change of the GW sound speed in the $\alpha_{\rm T}$ model can be explicitly expressed as
\al{
	{\cal S}(\eta ,K)
		={\cal S}_0(K\eta)+\alpha_{\rm T}^\ast\,\Delta{\cal S}(\eta ,K)\,\Theta (\eta -\eta_\ast)
	\,,
}
where ${\cal S}_0$ represents the standard part, which is the same as that in the standard cosmology with general relativity~\cite{Dai:2013kra}:
\al{
	{\cal S}_0(x)=\frac{1}{5}\int_0^x\dd\bar x\,\bar x\,{\cal T}_{h,0}(\bar x)\Biggl[ 1-\biggl(\frac{\bar x}{x}\biggr)^5\,\Biggr]
	\,,
}
and $\Delta{\cal S}$ is the correction part defined as
\al{
	\Delta{\cal S}(\eta ,K)
		=\frac{19}{20}(K\eta_\ast)^2\,{\cal T}_{h,0}(K\eta_\ast )\Biggl[ 1-\biggl(\frac{\eta_\ast}{\eta}\biggr)^5\,\Biggr]
			+\frac{1}{5}K^2\int_{\eta_\ast}^{\eta}\dd\bar\eta\,\bar\eta\,\Delta{\cal T}_h(\bar\eta ,K)
				\Biggl[ 1-\biggl(\frac{\bar\eta}{\eta}\biggr)^5\,\Biggr]
	\,.\label{eq:S in alpha_T model}
}

To see the behaviour of ${\cal S}$ in the $\alpha_{\rm T}$ model, in Fig.~\ref{fig:S} we plot the function ${\cal S}_0$
(black) and $\Delta{\cal S}$ for $K\eta_\ast =1$ (red), $2.7437$ (orange), $5$ (green), 
$10$ (blue), $30$ (purple), and $50$ (magenta) as a function of $K\eta$.
We find from Fig.~\ref{fig:S} that
not only the standard part ${\cal S}_0$ but also the correction part $\Delta{\cal S}$ persists after the tensor mode has decayed.
To understand the asymptotic behaviour of $\Delta{\cal S}$, taking 
the late-time limit while keeping $K\eta_\ast$ constant, we obtain
\al{
	\Delta{\cal S}(K\eta\to\infty )\to\frac{9}{4}K\eta_\ast\, j_1(K\eta_\ast )
	\,.
}
Moreover, the effect of the second-order contribution to the modified Poisson equation, namely $\Gamma_1$, on the tensor clustering fossils 
is much larger than other effects. This fact implies that to extract the information of the modified GW sound speed
in the context of the tensor clustering fossils, it is important to consider the dynamics of the scalar and tensor modes simultaneously.

%
\begin{figure}
\includegraphics[width=170mm]{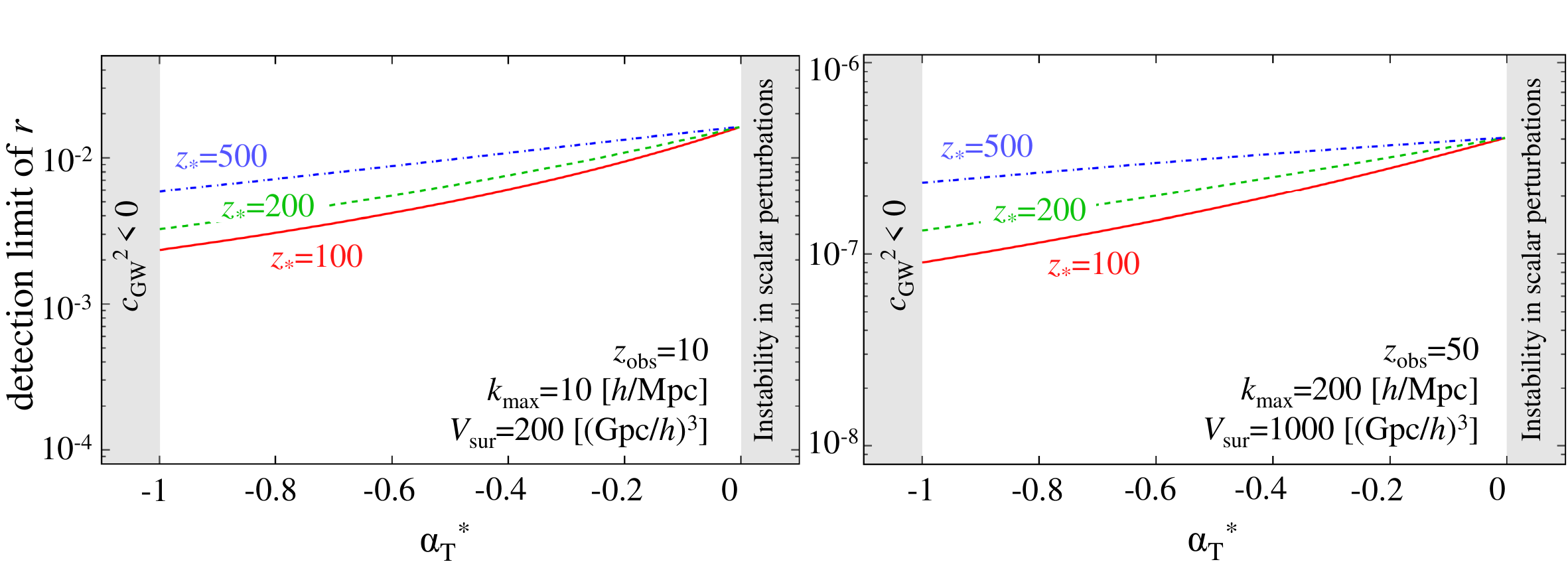}
\vspace{-2mm}
\caption{
The smallest value of the tensor-to-scalar ratio $r$ detectable at the 3$\sigma$ level as a function of
the amplitude of the instantaneous change in the GWs sound speed, $\alpha_{\rm T}^\ast$.
We show the results for survey parameters of $\{ z_{\rm obs}, k_{\rm max}, V_{\rm sur}\} =\{10,10\,[h/{\rm Mpc}],200\,[({\rm Gpc}/h)^3]\}$
and $\{50,200\,[h/{\rm Mpc}],1000\,[({\rm Gpc}/h)^3]\}$ with $z_\ast =100$ (red solid), $200$ (green dotted), and $500$ (blue dot-dashed), respectively.
}
\label{fig:rMin-alphaT}
\end{figure} 
%

Before closing this section, we would like to discuss the detectability of the instantaneous change 
in the sound speed of GWs.
To do this, we quantify the signal-to-noise ratio for the tensor power spectrum $P_h(K)$ estimated from the tensor clustering fossils as~\cite{Masui:2010cz}
\al{
	\frac{S}{N}=\Biggl[V_{\rm sur}\int\frac{\dd^3{\bm K}}{(2\pi)^3}\left(\frac{P_h(K)}{P_h(K)+N_h(K)}\right)^2\Biggr]^{1/2}
	\,,
}
where $V_{\rm sur}$ is the volume of the survey, and $N_h(K)$ is the tensor noise power spectrum.
In practice, the tensor mode can be reconstructed by applying the quadratic estimators to the density field.
Following \cite{Jeong:2012df}, we have the reconstructed tensor noise power spectrum $N_h$ as
\al{
	N_h(K)=\Biggl[\frac{1}{8}\int\frac{\dd^3{\bm k}}{(2\pi)^3}\frac{(1-\widehat{\bm K}\cdot\widehat{\bm k})^2P^2(k)}{2P^{\rm tot}(k)P^{\rm tot}(|{\bm K}-{\bm k}|)}\biggl\{\frac{1}{2}\frac{\dd\ln P}{\dd\ln k}+2{\cal S}_N(K)\biggr\}^2\Biggr]^{-1}
	\,,
}
where $P^{\rm tot}$ denotes the power spectrum of the matter field including noise.
We assume that the factor $\frac{1}{2}\frac{\dd\ln P}{\dd\ln k}+2{\cal S}_N$ can be approximately treated as a constant value
for large $k$ and introduce the largest wavenumber $k_{\rm max}$ for which the matter power spectrum can be measured
with high signal-to-noise, $P(k)/P^{\rm tot}(k)\approx 1$ for $k<k_{\rm max}$ and $P(k)/P^{\rm tot}(k)\approx 0$ for $k>k_{\rm max}$.
With these assumptions, the noise power spectrum reduces to \cite{Masui:2017fzw}
\al{
	N_h(K)\approx\frac{45(2\pi)^2}{k_{\rm max}^3}\biggl[\frac{1}{2}\frac{\dd\ln P}{\dd\ln k}+2{\cal S}_N(K)\biggr]^{-2}
	\,.
}
We adopt the standard inflationary form for the tensor power spectrum as $P_h(K)=2\pi^2 rA_{\rm s}/K^3$,
where $A_{\rm s}$ denotes the amplitude of the primordial scalar perturbation.
Once the survey parameters $\{z_{\rm obs},k_{\rm max},V_{\rm sur}\}$ and model parameters $\{\alpha_{\rm T}^\ast ,\eta_\ast (z_\ast)\}$ 
are specified, we can evaluate the significance of the detection by using these expressions.
We set $A_{\rm s}=2.12\times 10^{-9}$ and $\frac{\dd\ln P}{\dd\ln k}=-2.75$ as the fiducial model.
In Fig.~\ref{fig:rMin-alphaT}, we show the projected detection limit of the tensor-to-scalar ratio $r$ at the 3$\sigma$ level
as a function of the amplitude of the instantaneous change in GW sound speed, $\alpha_{\rm T}^\ast$,
for surveys with various parameters.
One can easily find that when $|\alpha_{\rm T}^\ast |$ is large, the detection limit of $r$ becomes small. 
Assuming the survey with $z_{\rm obs}=10$ and $V_{\rm sur}=200\,[({\rm Gpc}/h)^3]$ measuring scales down to $k_{\rm max}=10\,[h/{\rm Mpc}]$,
we find that if $r=10^{-2}$, $\alpha_{\rm T}^\ast <-0.17$, $-0.25$, and $-0.50$ for $z_\ast =100$, $200$, and $500$, respectively, could be detected.
For the futuristic survey that can be achieved by the 21-cm line survey of the dark ages, 
we consider the survey parameters $z_{\rm obs}=50$, $k_{\rm max}=200\,[h/{\rm Mpc}]$ and $V_{\rm sur}=1000\,[({\rm Gpc}/h)^3]$.
In this case, all the allowed parameter ranges of $\alpha_{\rm T}^\ast$ can be reached even if $r=5\times 10^{-7}$.

\section{Conclusion}
\label{sec:Conclustion}

We have studied the tensor clustering fossils as a possible probe of the theory of gravity, in particular 
the properties of large-scale tensor perturbations.
We have considered the effect of the modification of the theory of gravity on the tensor clustering fossils
and applied the resulting expressions to the model with the deviation of the sound speed of GWs from the speed of light.
Using the perturbative treatment of gravitational clustering and introducing the effective description of the Poisson equation,
which includes
the novel phenomenological model of the scalar-tensor tidal interactions, 
we have reformulated the tensor clustering fossils without assuming the explicit form of the gravity model [Eq.~\eqref{eq:calS}].
We found that the tensor clustering fossils can arise from the propagation of the large-scale tensor mode, 
the growth of the large-scale structure, and the scalar-tensor tidal interactions from the effective Poisson equation.
This fact implies that the tensor clustering fossils can be used to constrain the underlying theory of gravity in the late-time Universe.
To apply the derived expression, we considered the Horndeski scalar-tensor theory as the underlying gravity model, and
we have constructed the small-scale effective Lagrangian with the tidal interactions between the scalar and tensor modes 
in the language of the effective field theory of dark energy [Eq.~\eqref{eq:eff Lag}].
On the basis of the resulting effective Lagrangian, we have consistently 
solved the second-order equations of motion for the gravitational potentials
and have derived the formula applicable to the tensor clustering fossils in the Horndeski theory [Eqs.~\eqref{eq:mu}--\eqref{eq:Gamma2}].
Finally, we have applied the resulting formula to constrain the sound speed of GWs as a demonstration. 
Considering the specific model with the instantaneous change in the GW sound speed, we have discussed the detectability of the tensor clustering fossils 
by using the quadratic estimator of the density field.
We found that the measurement with the large survey volume, which can be achieved by the futuristic 21-cm line survey, can detect 
the large-scale GWs with the modified sound speed. 

In this paper, we have neglected several realistic effects such as the redshift-space distortion and the nonlinear growth of the 21-cm fluctuations.
In particular, the nonlinear growth of structure affects the small-scale 21-cm spectrum and the cutoff scale due to the nonlinear 
mode coupling should be taken into account~\cite{Yamauchi:2022fri}.
The large-scale tensor mode considered in this paper can also contribute to the intrinsic alignment of galaxy shape (see e.g. \cite{Schmidt:2012nw,Akitsu:2022lkl,Philcox:2023uor}).
It would be interesting to study the observational constraints on the properties of GWs through such observations. 
These are left to be investigated in future work.

\acknowledgements

We thank Tomoki Katayama for useful discussions.
This work was supported by
JSPS KAKENHI Grant Number 22K03627, 23K25868.


\end{document}